# The enhancement of superconducting transition temperature in yttrium sesquicarbide system, $Y_2C_3$, with the maximum $T_c$ of $T_c$=18K

Gaku AMANO, Satoshi AKUTAGAWA, Takahiro MURANAKA, Yuji ZENITANI and Jun AKIMITSU

*Department of Physics, Aoyama-Gakuin University, Fuchinobe 5-10-1, Sagamihara-shi, Kanagawa 229-8558*

We discovered the enhancement of superconducting transition temperature, $T_c$, in yttrium sesquicarbide system with the maximum $T_c$ of $T_c$=18K and their superconducting properties were discussed. The crystal structure of $Y_2C_3$ is the body-centered cubic ($Pu_2C_3$-type) structure, and the lattice parameters are varied with the heat treatment conditions. The magnetization ($M$-$H$) curves of this compound showed a typical type-II superconducting behavior, and the lower critical field, $H_{c1}(0)$, is 3.5mT. The $T_c$-value in this system was changed in the range from 15K to 18K, depending on the sintering conditions. In a previous report by Krupka *et al.*, maximum $T_c$ in $Y_2C_3$ was observed at 11.5K by the magnetic measurements. So we successfully synthesized the new high-$T_c$ phase in $Y_2C_3$.

The enthusiasm caused by the discovery of superconductivity in $MgB_2$[1] has yielded much research of boride compounds from various experimental and theoretical sides, and the discovery of $MgB_2$ also allows a search for a high-$T_c$ material in the similar system which contains light elements, B and C. So we concentrated our interest for searching a new material which shows high $T_c$, especially in lanthanoid carbide system. The interest for the superconducting lanthanoid carbides was increased due to the superconductivity with $T_c$ up to 23K in the Ni and Pd based borocarbide families[2–4] and also because of the observation of superconductivity with $T_c$ up to 11.6K in layered rare-earth carbide halides.[5,6]

Lanthanoid dicarbides $LnC_2$ ($Ln$: Y, La) crystallize in the body centered tetragonal (b.c.t.) $CaC_2$ structure type.[7] Superconductivity in the yttrium carbide system was first observed in the b.c.t. $YC_2$ at 3.88K,[8] furthermore, a strong dependence of the superconducting properties on the carbon deficiency was observed. Being stimulated by the interest for the layered yttrium carbide halide superconductors, of which crystal structures and chemical bonding properties are closely related to those of the lanthanoid dicarbides, the investigation of superconducting properties in $YC_2$, $Y_{1-x}Th_xC_2$ and $Y_{1-x}Ca_xC_2$ was carried out in detail.[9] In a previous report by Krupka *et al.*, the yttrium sesquicarbide phase crystallizing in the b.c.c. ($Pu_2C_3$-type) structure has been synthesized by arc melting and high pressure techniques.[10] Krupta *et al.* reported that high temperature annealing at ambient pressure of samples pre-pared by the high pressure (1.5-2.5GPa) techniques destroyed the $Pu_2C_3$-type structure. The magnetic susceptibility exhibited that the material was superconducting over the entire homogeneity range with $T_c$ varying from 6K to 11.5K. In a similar way, in the case of yttrium sesquicarbide, Th substituted sesquicarbides of yttrium and lantanum, $(Ln_{2-x}Th_x)C_3$, had attracted particular attention because their $T_c$ reached to the values close to those of niobium based A15-type compounds.[11] The phase, crystallizing in the b.c.c. ($Pu_2C_3$-type) structure, existed over a wide ternary range and the lattice parameters depend on the Y/Th metal and $C/M_T$ carbon-total metal atomic ratios. The material showed superconductivity with a variable $T_c$, having a maximum at 17K in nominal composition $Y_{0.7}Th_{0.3}C_{1.55}$ phase. So we tried to investigate the relationship between lattice parameters and superconducting properties in $Y_2C_3$, and synthesize this material under higher temperature and pressure conditions. In general, high-pressure synthesis is a very efficient method of the search for new superconducting materials.[12] A closed system in high-pressure synthesis is effective not only for stabilizing a composition but also for expanding a solid-solution range. As for carbides, it is difficult to prepare a highly dense polycrystalline sample when using a conventional solid-state reaction method because the melting temperature is extremely high.

In this study, we present the superconductivity of yttrium sesquicarbide $Y_2C_3$ phase prepared under high pressure and discussed the compositional dependence of crystal structure and superconductivity.

The mixing powders with the appropriate amounts of powders Y (99.9%) and C (99.95%) in a dry box were mounted into a BN cell. The samples were heated to 1600 °C in a few minutes and kept for 10-30 minutes under the pressure of 4-5.5GPa using a cubic-anvil-type equipment. They were quenched to room temperature within a few seconds. The polycrystalline samples were examined by powder X-ray diffraction, being performed by a conventional X-ray spectrometer with a graphite monochromator (RINT-1100 RIGAKU) and the intensity data were collected with CuK$\alpha$ radiation over a $2\theta$ range from 5° to 80° at 0.02° step width. The magnetic susceptibility and magnetization measurements were performed with the SQUID magnetometer (MPMSR2 Quantum Design Co. Ltd.) and the PPMS system (Quantum Design Co. Ltd.).

Figure 1 shows the powder X-ray diffraction pattern of $Y_2C_3$, including impurity phases, C, $Y_2O_3$, $YC_2$, $Y_2O_xC_y$ etc. The main phase is indexed as a cubic unit cell with space group: $I\bar{4}3d$. Especially, we confirmed that $Y_2O_xC_y$ phase did not show superconductivity by themselves. The crystal structure of $Y_2C_3$ is shown in Fig. 2. The lattice parameter, $a$, calculated from all indexes is 8.226Å. However, the lattice parameters of $Y_2C_3$ phase change for the different sintering process in sample preparation and exist in the range between 8.181Å and 8.226Å, and they are not in agreement with the previous data.[10] In a previous report, the lattice parameters exist in the range between 8.2142Å±5 and 8.2514Å±12,[10] and these parameters are rather larger than our results. We considered that this behavior is originated from the difference of pressure range (this work; 4-5.5GPa and Krupta *et al.*; 1.5-2.5GPa) in the sample preparation.

The superconductivity of $Y_2C_3$ was confirmed by DC susceptibility in a magnetic field of 10 Oe, as shown in Fig. 3. The magnetic susceptibility of $Y_2C_3$ exhibits remarkable drop at 18K, suggesting the occurrence of superconductivity and the volume fraction of superconductivity at 5K is estimated to be about 10%. Moreover, the clear diamagnetic signals are observed at the range between 15K and 18K (Fig. 4) for the samples synthesized under several conditions. The $T_c$'s of samples exhibit remarkable change, depending on the heat treatment condition.

Figure 5 shows the magnetization versus magnetic field ($M$-$H$) curves, exhibiting the behavior of the typical type-II superconductor. From $M$-$H$ curves, we tried to estimate the lower critical field, $H_{c1}(T)$, being determined from the magnetic field, where the initial slope meets the extrapolation curve of $(M_{up}+M_{down})/2$ and these values are plotted as a function of temperature. The roughly estimated $H_{c1}(0)$ of $Y_2C_3$ are ∼3.5mT. The penetration depth, $\lambda(0)$, is calculated from the relationship between $H_{c1}(0)$ and $\lambda(0)$, $\mu_0 H_{c1} \sim \phi_0/\pi\lambda^2$, where $\mu_0$ and $\phi_0$ are magnetic permeability of vacuum and quantum flux, respectively, and the value $\lambda(0)$ can be obtained to be about 4340Å.

Figure 6 shows $T_c$ vs. lattice parameters in the present work as well as reference by Krupka *at al*. It is suggested that C contents in $Y_2C_3$ are varied by the heat treatment condition, and are also strongly related with the superconductivity in this system.

In conclusion, we discovered the enhancement of superconducting transition temperature, $T_c$, in yttrium sesquicarbide system with the maximum $T_c$ of $T_c$=18K and investigated their superconducting properties. The lattice parameters, $a$, obtained in this work exist in the range between 8.1805Å and 8.2261Å, and they are smaller than the values previously reported (8.2142-8.2514Å). Moreover the $T_c$'s in this system are varied from 15K to 18K and depends on the lattice parameters. The magnetization ($M$-$H$) curves of these compounds showed a typical type-II superconducting behavior, and the lower critical field, $H_{c1}(0)$, is 3.5mT. We successfully synthesized the new high-$T_c$ phase in $Y_2C_3$.

This work was partially supported by a Grant-in-Aid for Science Research from the Ministry of education, Science, Sports and Culture, Japan.

**References**

1) J. Nagamatsu, N. Nakagawa, T. Muranaka, Y. Zenitani and J. Akimitsu, Nature **410** (2001) 63
2) R. Nagarajan, C. Mazumdar, Z. Hossain, S.K. Dhar, K.V. Gopalakrishnan, L.C. Gupta, C. Godart, B.D. Padalia, and R. Vijayaraghavan, Phys. Rev. Lett. **72** (1994) 274
3) R.J. Cava, H. Takagi, B. Batlogg, H.W. Zandbergen, J.J. Krajewski, W.F. Peck, Jr., R.B. van Dover, R.J. Felder, T. Siegrist, K. Mizushashi, J.O. Lee, H. Eisaki, S.A. Carter, and S. Uchida, Nature **367** (1994) 146
4) R. J. Cava, H. Takagi, H.W. Zandbergen, J.J. Krajewski, W.F. Peck, Jr., T. Siegrist, B. Batlogg, R.B. van Dover, R.J. Felder, K. Mizuhashi, J.O. Lee, H. Eisaki, and S. Uchida, Nature **367** (1994) 252
5) A. Simon, A. Yoshiasa, M. Backer, R.W. Henn, R.K. Kremer, Hj. Mattausch, and C. Felser, Z. Anorg. Allg. Chem. **622** (1996) 123
6) R.W. Henn, W. Schnelle, R.K. Kremer, and A. Simon, Phys. Rev. Lett. **77** (1996) 374
7) M. Atoji, J. Chem. Phys. **35/6** (1968) 1950
8) A.L. Giorgi, E.G. Szklarz, M.C. Krupka, T.C. Wallace and N.H. Krikorian, J. Less-Common Met. **14** (1968) 247
9) Th. Gulden, R.W. Henn, O. Jepsen, R.K. Kremer, W. Schnelle, A. Simon, and C. Felser, Phys. Rev. B **56** (1997) 9021
10) M.C. Krupka, A.L. Giorgi, N.H. Krikorian and E.G. Szklarz, J. Less-Common Met. **17** (1969) 91
11) M.C. Krupka, A.L. Giorgi, N.H. Krikorian and E.G. Szklarz, J. Less-Common Met. **19** (1969) 113
12) For example, X. Cgen, T. Koiwasaki and S. Yamanaka, J. Solid State Chem. **159** (2001) 80

**Figure captions**

Figure 1: The powder X-ray diffraction pattern of $Y_2C_3$.

Figure 2: The crystal structure of $Y_2C_3$. The dark gray circles show Y atoms and the light gray circles show C atoms. The black lines in this figure show unit cell.

Figure 3: The temperature dependence of susceptibility in $Y_2C_3$.

Figure 4: The temperature dependence of susceptibilities in $Y_2C_3$ synthesized under several conditions. The inset shows the expansion near $T_c$ region.

Figure 5: The magnetization versus magnetic field ($M$-$H$) curves of $Y_2C_3$. Inset shows the $M$-$H$ curves at low field region.

Figure 6: The variation of $T_c$ with lattice parameter, $a$, in $Y_2C_3$ (the data observed in present work and reference by Krupka *et al.*)

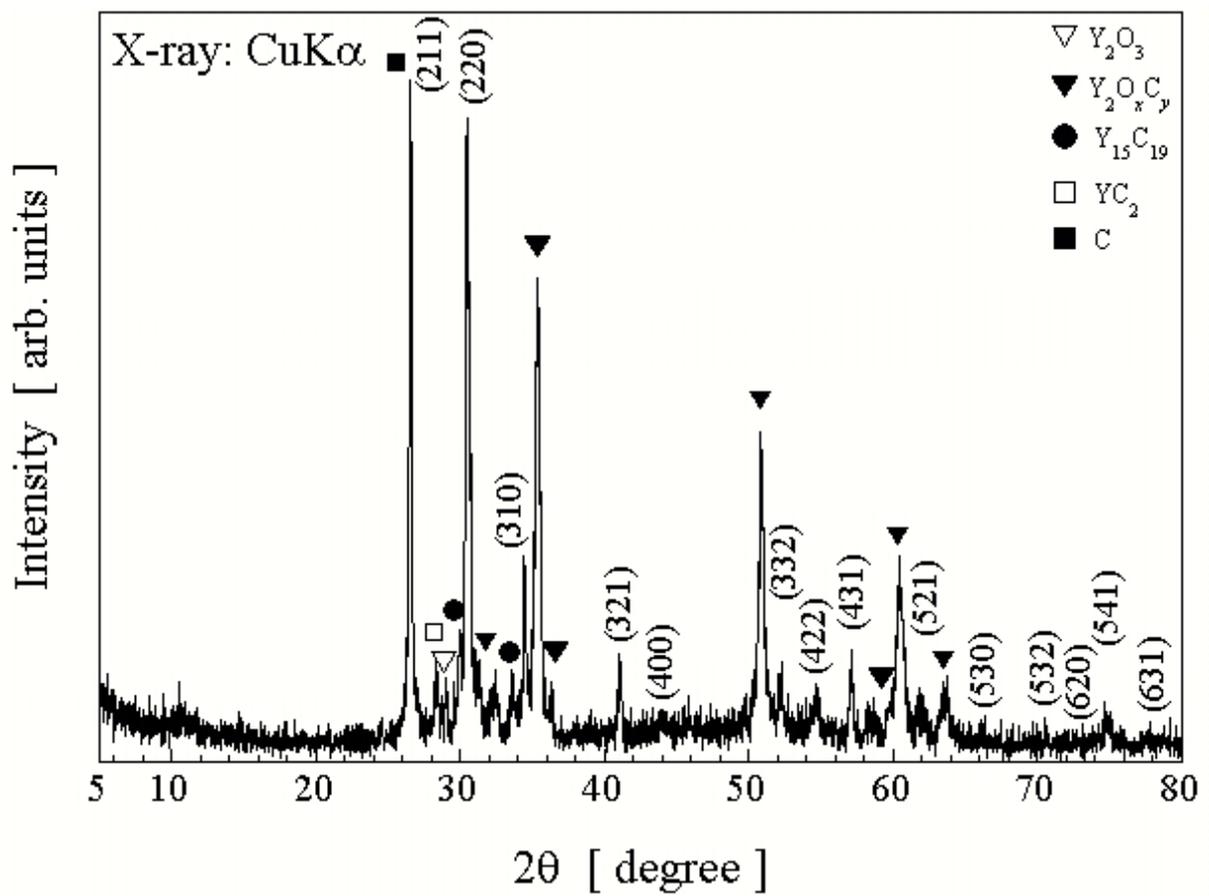

Fig. 1. The powder X-ray diffraction pattern of $Y_2C_3$.

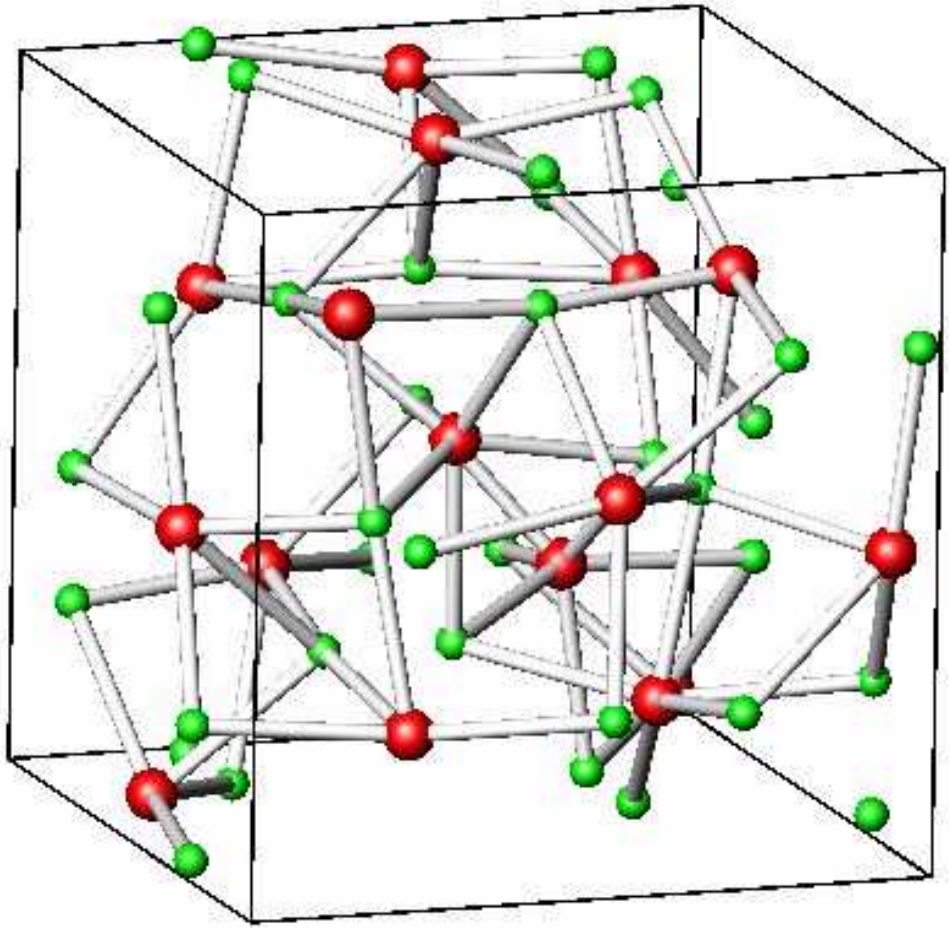

Fig. 2. The crystal structure of $Y_2C_3$. The dark gray circles show Y atoms and the light gray circles show C atoms. The black lines show unit cell.

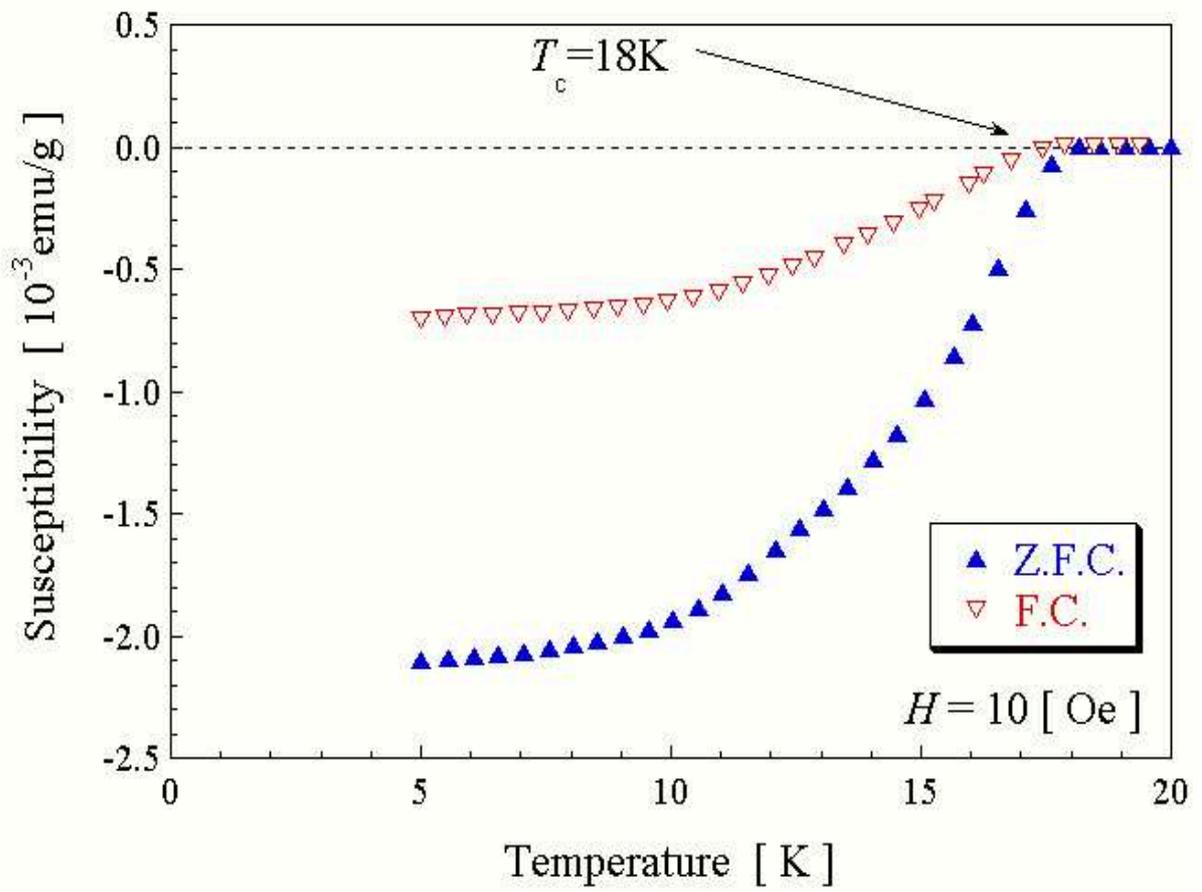

Fig. 3. The temperature dependence of susceptibility in $Y_2C_3$.

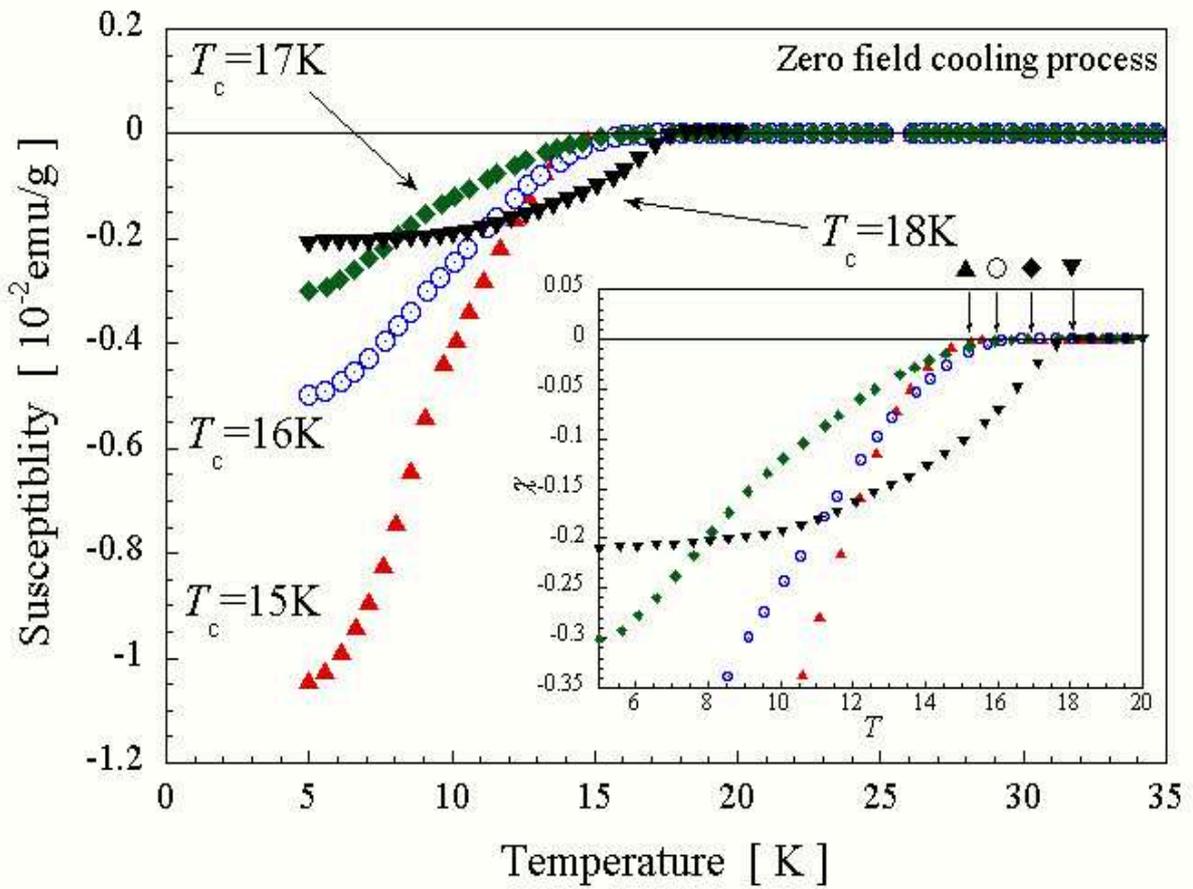

Fig. 4. The temperature dependence of susceptibilities in $Y_2C_3$ synthesized under several conditions. The inset shows the expansion near $T_c$ region.

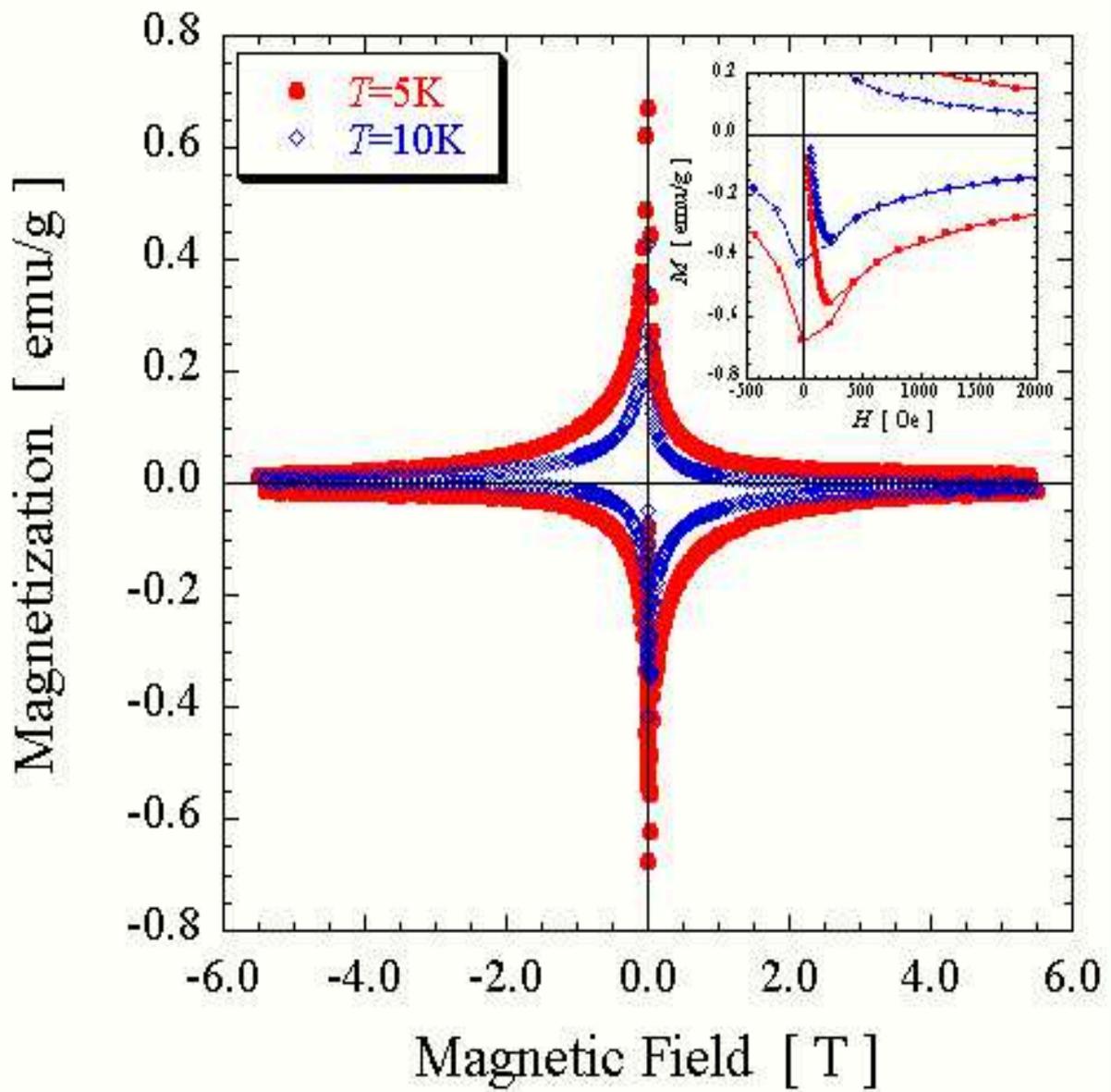

Fig. 5. The magnetization versus magnetic field ($M$-$H$) curves of $Y_2C_3$. Inset shows the $M$-$H$ curves at low field region.

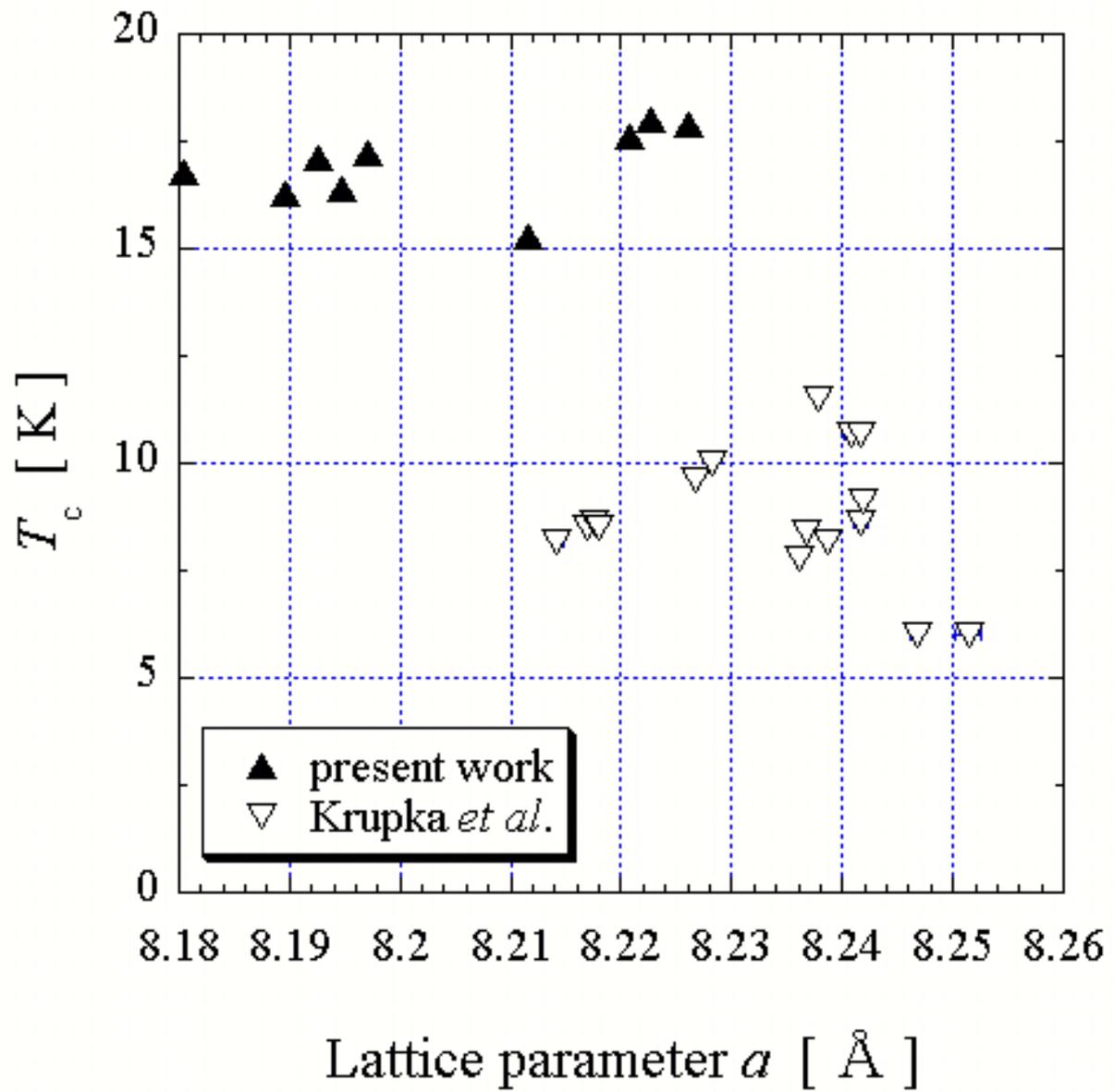

Fig. 6. The variation of $T_c$ with lattice parameter, $a$, in $Y_2C_3$